\documentclass[Royal]{sagej}

\usepackage{moreverb,url}
\usepackage[colorlinks,bookmarksopen,bookmarksnumbered,citecolor=blue,urlcolor=blue]{hyperref}
\usepackage{tikz-opm}
\usepackage{amsmath, amsthm, amscd, amsfonts,graphicx}
\newtheorem{theorem}{Theorem}[]

\newtheorem{proposition}[theorem]{Proposition}

\theoremstyle{definition}
\newtheorem{definition}[theorem]{Definition}

\newcommand\BibTeX{{\rmfamily B\kern-.05em \textsc{i\kern-.025em b}\kern-.08em
T\kern-.1667em\lower.7ex\hbox{E}\kern-.125emX}}

\begin{document}


\title{Hyper-rational choice theory}

\author{Madjid Eshaghi Gordji\affilnum{1} and Gholamreza Askari\affilnum{1}}

\affiliation{\affilnum{1}Department of Mathematics, Semnan University P.O. Box 35195-363, Semnan, Iran}


\email{meshaghi@semnan.ac.ir; g.askari@semnan.ac.ir}

\begin{abstract}
The rational choice theory is based on this idea that people rationally pursue goals for increasing their personal interests. In most conditions, the behavior of an actor is not independent of the person and others' behavior. Here, we present a new concept of rational choice as a \emph{hyper-rational choice} which in this concept, the actor thinks about profit or loss of other actors in addition to his personal profit or loss and then will choose an action which is desirable to him. We implement the hyper-rational choice to generalize and expand the game theory.  Results of this study will help to model the behavior of people considering environmental conditions, the kind of behavior interactive, valuation system of itself and others and system of beliefs and internal values of societies. Hyper-rationality helps us understand how human decision makers behave in interactive decisions.
\end{abstract}


\maketitle

\section*{Introduction}
The rational choice theory is one of the theories which predict humans and societies' behavior in social sciences. The main hypothesis in the prediction of humans' behavior is that these behaviors have almost rational and targeted trend. In this theory, society is a set of the people which has rational action. Principles and fundamentals of the rational actor are the basis of all social sciences theories which have been explained by some theorists such as Marx Weber \cite{weber1978economy}, James Coleman \cite{gamst1991foundations}, Raymond Boudon \cite{boudon1998limitations}, John Elster \cite{elster1989cement} this theory has been also criticized by many theorists. Harsanyi \cite{harsanyi1980advances} defines rational action in routine life as a behavior through which people choose the best accessible means to achieve a definite goal. The first and yet the main application of the rational choice theory beyond economics is to use it in politics \cite{udehn2002methodological}. Herbert Simon is most famous for what is known to economists as the theory of bounded rationality, bounded rationality is the idea that in decision-making, rationality of individuals is limited by the information they have, the cognitive limitations of their minds, and the finite amount of time they have to make a decision \cite{simon1955behavioral}. The most important theories based on the rational choice theory is decision-making theory \cite{raiffa1974applied, abelson1985decision}, Collective action theory \cite{olson1965logic} and game theory \cite{von2007theory}. The theory of rational choice is one of the most important components of many game theory models. Hence, the rationality of the players is one of the fundamental principles which led to generalization and expansion of this theory. The assumption of rational behavior is based on the fact that each player intends to optimize or seek the maximum benefit.

A choice of behavior in a single person decision-making problem is called player's action. The set of actions available is shown with $A$. Function $u:A\to R$ which relates a outcome such as $u(a)$ with numerical value to every member $a$ in $A$ is called payoff function. An individual’s desires lead to a ranking of outcomes in terms of preference \cite{webb2007game}. If a person prefers outcome $a_1$ to outcome $a_2$, we will show it as $a_1$ > $a_2$. If a person is indifferent between the two outcomes $a_1$ and $a_2$, it will be shown as $a_1 \sim a_1$. Showing $a_1\succeq a_2$ means that person either prefers outcome $a_1$ to outcome $a_2$ or is indifferent between the two outcomes.

\begin{definition}(rationality )\label{D.1}
An individual will be called \textit{rational} under certainty if their preferences for outcomes satisfy the following conditions:
\begin{enumerate}
  \item (Completeness) Either $a_1\succeq a_2$ or $a_2\succeq a_1$
  \item (Transitivity) If $a_1\succeq a_2$ and $a_2\succeq a_3$ then $a_1\succeq a_3$
\end{enumerate}
\end{definition}

The completeness condition ensures that all outcomes can be compared with each other. The transitivity condition implies that outcomes can be listed
in order of preference \cite{webb2007game}. One of the important problems of the rational choice theory is that it does not describe the rational behavior of human subjects as well. In the previous methods, failure to pay attention to outcome obtained for other decision makers is the main negligence which can have a considerable effect on the interactive decisions. We are looking to present a new concept of rationality, that can serve as a new foundation for human thinking.

This study demonstrates that how human decision makers behave in interactive decisions. Here, in addition to the individual profit which has been mentioned in the rational choice theory, we focused on profit or loss of other players which led to the division of persons' preferences into two classes.  The first class is the individual preferences to maximize his benefit or minimize his loss and the second class is preferences of the individual to maximize profit or loss of others (The person's preferences for others). The individual's choice among his preferences: individual preferences, preference for others and both of them at the same time, makes his hyper-preferences. The advantage of this concept emphasizes the importance of outcomes of other actors in the game. Based on the idea, in most conditions, the hyper-rationality behavior of an actor is dependent on the person and others’ behavior. Therefore, we present the unified approaches to analysis and cognition of humans and societies’ behavior in social sciences. We implement the hyper-rational choice for interpretation strategic interaction between players. For this purpose, we will examine Prisoner’s Dilemma, Missile crisis game and Trickery game.

\section*{Results}
\subsection*{Hyper-rational Choice}
Numerous results from experimental economics have shown that classic rationality assumptions do not describe the behavior of real human subjects. For example, it is not uncommon for people, in experimental situations, to indicate that they prefer $A$ to $B$, $B$ to $C$, and $C$ to $A$ \cite{alexander2010evolutionary}. If we consider profit or loss of other persons given rational behaviour of the persons which is based on individual benefit, some of the human behaviors may be described. In 1926 Ragnar Frisch developed for the first time a mathematical model of preferences in the context of economic demand and utility functions \cite{barten1982consumer, arrow1991handbook}. In 1944, Von Neumann and Morgenstern introduced the preferences as a formal relation. \cite{von2007theory}.

Here we divide preferences of people into two classes. The first class is a set of \textit{individual preferences} (include: individual profit, individual loss and indifferent between profit and loss) of the person which seek to maximize his profit or minimize his loss. The second class is a set of \textit{preferences for others} (include: profit for other, loss for other and indifferent between profit or loss for others) of the person which seek to maximize profit or loss of other actors. Based on the second class of preferences (The person's preferences for others), assumption about profit or loss of other actors can be added in the form of human behaviors such as altruism, devotion, fraud, jealousy, and mistrust to the concept of rationality. As a result, the person has three choices about his preferences: individual preferences, preferences for other actors and both of them simultaneously. Therefore, an individual’s desires in a multiple person decision–making the problem that leads to a ranking of preferences in terms of \textit{hyper-preferences}. Hence, the hyper-preferences for each individual specifies six behavioral options. Here, we consider four main behavioral options: individual profit, individual loss, profit for others and loss for others. In the following, we want to turn a rational individual into a hyper-rational individual.

Now consider a rational individual. The set of possible choices of rational individual $i \in\{1,2,..., n\}$ is shown with $A_i=\{a_1, a_2, ..., a_n\}$. Given hyper-preferences, how will a hyper-rational individual behave? We assume that given a set of choices $B\subseteq\mathcal{A}=A_1\times A_2 \times ... \times A_n$. We define the weak hyper-preferences of actor $i$' over the set $B$ as follows:
\begin{align*}
&(a_1, a_2,...,a_n)_i\succeq^{'}(b_1, b_2,..., b_n)_i \Leftrightarrow either~a_1\succeq b_1~or~a_1\preceq b_1~based~on~actor~i'~preferences\\
&\;\;\;\;\;\;\;\;\;\;\;\;\;\;\;\;\ ~for~actor~1~and~either~a_2\succeq b_2 ~ or~ a_2\preceq b_2 ~based~on~actor~i'~preferences\\
&\;\;\;\;\;\;\;\;\;\;\;\;\;\;\;\;\ for~actor~2~and~either~a_i\succeq b_i~or~a_i\preceq b_i~ based~on~actor~i'~preferences\\
&\;\;\;\;\;\;\;\;\;\;\;\;\;\;\;\;\ and~either~a_n\succeq b_n~or~a_n\preceq b_n~ based~on~actor~i'~preferences~for~actor~n,
\end{align*}
where relation $\succeq$ is complete and transitive. We say that $(a_1, a_2,...,a_n)$ is strictly preferred to $(b_1, b_2,..., b_n)$, or $(a_1, a_2,...,a_n) \succ^{'} (b_1, b_2,..., b_n)$, if $(a_1, a_2,...,a_n) \succeq^{'} (b_1, b_2,..., b_n)$ but not $(b_1, b_2,..., b_n) \succeq^{'} (a_1, a_2,...,a_n)$. We say the actor is indifferent between $(a_1, a_2,...,a_n)$ and $(b_1, b_2,..., b_n)$, or $(a_1, a_2,...,a_n) \sim^{'} (b_1, b_2,..., b_n)$, if $(a_1, a_2,...,a_n) \succeq^{'} (b_1, b_2,..., b_n)$ and $(b_1, b_2,..., b_n) \succeq^{'} (a_1, a_2,...,a_n)$. So, we defined set of hyper-preference over set of preferences.

\begin{definition}
The relation $\succeq^{'}$ on $B$ is \emph{complete} if for all $(a_1, a_2,...,a_n), (b_1, b_2,..., b_n)\in B$ either $(a_1, a_2,...,a_n) \succeq^{'} (b_1, b_2,..., b_n)$ or $(b_1, b_2,..., b_n) \succeq^{'} (a_1, a_2,...,a_n)$, or both.
\end{definition}

The completeness condition ensures that all action profiles can be compared with each other.  Hyper-preferences help the person to consider more realistic conditions for decision-making and specify his objectives. Hyper-preferences of a individual has \textit{taxonomy} characteristic. Taxonomy of hyper-preference means that if we face an actor with two choices of hyper-preferences, she will necessarily have an opinion on which she likes more. Taxonomy of actor's hyper-preferences depends on environmental condition, the kind of behavior interactive (kind of interactive person), beliefs system of itself and others and valuation system of itself and others. Taxonomy of hyper-preferences helps to a person to retrieve his/her preferences and reduces the rate of computation. Taxonomy of hyper-preferences helps the actor checked the situation rapidly and the decision-making process and its criteria are discovered. Taxonomy of hyper-preferences helps to individual to consider benefit of observing the value of norms or loss resulting from not observing them. Considering situations which the individual faces, we define hyper-rationality as follows:

\begin{definition}(Hyper-rational )
An individual will be called \textit{hyper-rational} under certainty if is a rational (see Definition $\ref{D.1}.$) and their hyper-preferences for preferences (individual or for others) satisfy at least one of the following conditions:
 \begin{enumerate}
\item The actor chooses from the set of available alternatives (actions) based on individual preferences;
\item The actor chooses from the set of available alternatives (actions) based on preferences for other actors.
\end{enumerate}
\end{definition}

It can be concluded that each hyper-rational actor is a rational actor, but each rational actor is not a hyper-rational actor. In this concept, we call a person hyper-rational if he considers profit or loss of others in his interaction with other people in society in addition to individual profit or loss. Based on the concept of hyper-rationality, an actor may not identify an action which has the maximum benefit for him but may be able to choose an action which has the maximum profit or loss for other actors. So, attention to the emphasis on the importance of the outcome of other actors in the game is among the advantage of this concept. Hyper-rationality doesn’t simplify decision-making but specifies how to make a decision. This concept promotes both cooperative and non-cooperative behavior, But in practice, more cooperation is considered. Hyper-rational actor removes information cognitive and processing constraints based on the taxonomy of hyper-preferences. The hyper-rational actor makes a decision based on his hyper-preferences and chooses an action which is based on his targets despite knowing no access to full information. The number of decision-making stages is different among persons considering prioritization of hyper-preferences; therefore, hyper-rational doesn't mean limited searching. In definition of hyper-rationality, condition $(2)$ from Definition $\ref{D.1}.$ is not necessary. In other words, if an individual simultaneously considers two classes of preferences, transitivity property of outcomes may do not exist. For example, in Prisoner’s Dilemma based on collective profit or collective loss, transitivity property hyper-preferences do not exist.

Some human behaviors can be expressed with help of hyper-rationality concept. The emergence of behavior such as jealousy in a person is due to the fact that since the person cannot increase his benefit and is unhappy of that, he prefers not to give more benefit to others. Also, a benevolent player can choose an action based on his hyper-rationality which will bring more benefits to others. In contrast to a rational player with a hyper-rational player, the hyper-rational player has more chance of winning. Therefore, the concept of hyper-rationality can give a new concept of rationality called \textit{hyper-rational choice theory}. Based on the new concept of hyper-rational choice, some of the human behaviors can be modeled with help of game theory.


\section*{Hyper-rational Choice as the basis for game theory}
Game theory aims to help us understand situations in which decision-makers interact. Like other sciences, game theory consists of a collection of models. A model is an abstraction we use to understand our observations and experiences \cite{osborne2004introduction}. The concept of "game" indicates interactions between the players well, but this concept has not been clarified in the mathematical modeling. In other words, mathematical modeling is defective and cannot express dependency of strategy choice by a player on strategy choice by other players. In the mathematical modeling, the dependency of decision choice by players has been expressed only based on the individual benefit of the players and no discussion has been made about player's preferences of loss or profit of other players. In the theory of the game, different methods are used for finding equilibrium with help of rationality concept and one of these methods is the best response function \cite{mccarty2007political}. Based on the definition of the best response function, equilibrium is a point in which strategy chosen by each individual is the best response to the strategies chosen by other players and maximize personal profit \cite{harsanyi1986rational}.

Now, we apply hyper-rational choice theory as a basis and main element of modeling in game theory. With help of hyper-rationality, we analyze conditions of a strategic game. Hyper-rationality in game theory helps the player choose successful strategies of the game in interactive conditions and reproduce them. The concept of hyper-rationality helps the game theory enter other fields of sciences with more logical power. We assume that each player in the game is hyper-rational. Thus, a hyper-rational player will renormalize her opinion based on the common knowledge that each player is hyper-rational. Below, we show the best response functions based on hyper-preferences of players with  $B$, $P$, and $L$.  Precisely, we define the set-valued function $B_i$ by
\begin{align*}
B_i(a_{-i}):=\Big\{a_i\in A_i:u_{i}(a_{i}, a_{-i})\geq u_{i}(a^{'}_{i}, a_{-i})~for~all~a^{'}_{i}\in A_{i} \Big\},
\end{align*}
any action in $B_i(a_{-i})$ is at least as good based on individual benefit for player $i$ as every other action of player $i$ when the other players’ actions are given by $a_{-i}$. We call $B_i$ the best response function of player $i$ based on individual benefit. Precisely, we define the set-valued function $K_i$ by
\begin{align*}
P_i(a_{-i}):=\Big\{a_i\in A_i:u_{-i}(a_{i}, a_{-i})\geq u_{-i}(a^{'}_{i}, a_{-i})~for~all~a^{'}_{i}\in A_{i} \Big\},
\end{align*}
any action in $K_i(a_{-i})$ for player $i$ relative to every other action of player $i$ is at least the best based on profit for other players when the other players’ actions are given by $a_{-i}$. We call $K_i$ the best response function of player $i$ based on profit for other players. Precisely, we define the set-valued function $L_i$ by
\begin{align*}
L_i(a_{-i}):=\Big\{a_i\in A_i:u_{-i}(a_{i}, a_{-i})\leq u_{-i}(a^{'}_{i}, a_{-i})~for~all~a^{'}_{i}\in A_{i} \Big\},
\end{align*}
any action in $L_i(a_{-i})$ for player $i$ relative to every other action of player $i$ is at least as good based on the loss for other players when the other players’ actions are given by $a_{-i}$. We call $L_i$ the best response function of player $i$ based on the loss of other players. In competitive interactions, we define strictly dominant action and weakly dominant action based on the loss of other players.

\begin{definition}( Strict domination of loss)
In a strategic game for player $i$, action $a^{''}_{i}$ is strictly dominant on her action $a^{'}_{i}$ for loss of others, if we have
\begin{equation*}
u_{-i}(a^{''}_{i}, a_{-i})< u_{-i}(a^{'}_{i}, a_{-i}) \;\;\;\;\ for\; every \;\;\;\ a_{-i}\in A_{-i},
\end{equation*}
\end{definition}
where $u_i$ is a payoff function that represents player $i$’s preferences. It is defined as strictly dominant action based on benefit for other players similar, but the difference is that direction of the relation $<$ is changed.

\begin{definition}( Weak domination of loss)
 In a strategic game for player $i$, action $a^{''}_{i}$ is weakly dominant on her action $a^{'}_{i}$ for loss of others, if we have :
\begin{align*}
u_{-i}(a^{''}_{i}, a_{-i})\leq u_{-i}(a^{'}_{i}, a_{-i}) \;\;\;\;\ for\; every \;\;\;\ a_{-i}\in A_{-i}
\end{align*}
and
\begin{align*}
u_{-i}(a^{''}_{i}, a_{-i})< u_{-i}(a^{'}_{i}, a_{-i}) \;\;\;\;\ for\; some \;\;\;\ a_{-i}\in A_{-i},
\end{align*}
\end{definition}
where $u_i$ is a payoff function that represents player $i$’s preferences. It is defined as weakly dominant action based on benefit for other players similar, but the difference is that direction of relations $\leq$ and $<$ is changed.

The actions which are chosen based on the concept of hyper-rationality (hyper-preferences) and rationality of the players may be similar or different. To prevent ambiguity in interactions, we divide actions of players into three classes as strictly dominant action and weakly dominant action based on individual benefit, strictly dominant action and weakly dominant action based on  profit for other players and strictly dominant action and weakly dominant action based on the loss for others. The following proposition shows a method for finding equilibrium in the game.

\begin{proposition}
 The action profile $a^*$ is a equilibrium point of strategic game if and only if hold true in at least one of the following conditions:
\begin{itemize}
\item Each action of the player is the best response to actions of other players based on personal benefit:
\begin{align*}
a^*\;\; is \;\; in\;\; B_i(a^*_{-i}) \;\; for\;\; every\;\; player\;\; i,
\end{align*}
\item Each action of the player is the best response to actions of other players based on the benefit of other players:
\begin{align*}
a^*\;\; is \;\; in\;\; K_i(a^*_{-i}) \;\; for\;\; every\;\; player\;\; i.
\end{align*}
\item Each action of the player is the best response to actions of other players based on loss of other players:
\begin{align*}
a^*\;\; is \;\; in\;\; L_i(a^*_{-i}) \;\; for\;\; every\;\; player\;\; i.
\end{align*}
\end{itemize}
\end{proposition}

We consider equilibrium based on concept of Nash \cite{nash1950equilibrium, nash1951non}. Based on the concept of hyper-rationality, equilibriums can be divided into three classes. The first class is the equilibria which are considered based on personal benefit. The second class is the equilibria which are selected based on profit or loss of other players. The third class is the equilibria which are considered based on individual benefit and loss or profit of other players at the same time. An action profile may be selected which is the equilibrium point of the game based on hyper-preference of the maximum loss for other players and also has the maximum loss for all players, but it is not Nash equilibrium based on the classic concept of rationality. In addition, based on this concept, games may have one (second best), two (Prisoner's Dilemma), three (Trickery game) and four (Chicken game) equilibrium or may has not equilibrium. For example, the Matching pennies is not in equilibrium and Missile crisis game has two equilibrium. Hyper-rationality helps the analyst to interpret every cell of the game table and have more accurate analysis. Furthermore, with help of concept of hyper-rationality, we introduce a new definition of equilibrium in the game as \textit{hyper-equilibrium}.

\begin{definition}(Hyper-equilibrium )
The action profile $b^*=(a^{'}_i, a^*_{-i})$ in a strategic game is a hyper-equilibrium if , only for every player $i$ and every action $a^*_i$ of player $i$, $b^*$ is at least as good according to player $i$’s hyper-preferences as the action profile $(a^*_i, a^*_{-i})$ in which player $i$ chooses $a^*_i$
while every other players chooses $a^*_{-i}$. Equivalently, for every player $i$, based on benefit of other players
\begin{align*}
 u_{-i}(a^{'}_i, a^*_{-i})\geq u_{-i}(a^*_{i}, a^*_{-i}) \;\;\ for\; all \;\;\ a^{*}_{i}\in A_{i},
\end{align*}
or  for every player $i$, based on loss of other players
\begin{align*}
 u_{-i}(a^{'}_i, a^*_{-i})\leq u_{-i}(a^*_{i}, a^*_{-i}) \;\;\ for\; all \;\;\ a^{*}_{i}\in A_{i},
\end{align*}
\end{definition}
where $u_i$ is a payoff function that represents player $i$’s preferences. Hyper-equilibrium is a point in which only one player can increase or decrease outcome of other players by changing his action without other players having motivation for the change of action. In the next section, we explain how human decision makers behave in interactive decisions and also presented an important game which has special equilibrium which we call hyper-equilibrium.


\section*{Behavior of individuals in human societies}
In a social dilemma game, there are different behavioral options, for cooperation and competition \cite{hertel1994affective}. There are several studies which provide the main theoretical insights of the general psychological game framework \cite{geanakoplos1989psychological, attanasi2008survey}. The classical theory of game relies on assumptions of perfect rationality and full common knowledge that are far removed from the cognitive capacities of human players and of limited use in explaining human strategic behavior. On the other hand, blind experimentation is also unhelpful, because the proper use of experiments is to test hypotheses, and without good hypotheses, no useful progress can be made \cite{colman2015psychology}. Interactions between psychologists and economists have been marked more by conflict than by collaboration and the absence of a common research language impedes communication between this the disciplines\cite{manski2017collaboration}. Some of the studies describe the relationship between psychologists and economists \cite{croson2004explaining, attanasi2008survey, schotter2014belief, de2017eliciting}. In this section, we review and interpret Prisoner’s Dilemma, Missile crisis game and Trickery game, with the help of the concept of hyper-rationality.

\subsection*{Prisoner's Dilemma}
The essence of cooperation is described by the Prisoner’s Dilemma \cite{dreber2008winners}. Cooperation is a hallmark of human society \cite{trivers1971evolution, hamilton1981evolution, rockenbach2006efficient,  nowak2006five, rand2009positive, rand2011dynamic, rand2012spontaneous, rand2014static}. In theory of games, there has been much interest in exploring the Prisoner’s Dilemma. Here we investigate depends on the claim that each player in the Prisoner’s Dilemma is hyper-rational. In Fig. \ref{fig:PC} based on definitions of hyper-rationality, in Prisoner's Dilemma $g_1$ for both players based on loss of other players, defect $D$ is a strictly dominant action of loss and cooperate $C$ is a strictly dominated action of loss. In this game, equilibrium point $(D, D)$ is an equilibrium which the players selected based on loss of the opponent. For both players, based on the benefit of other players, Cooperate is a strictly dominant action of profit and defect is a strictly dominated action for profit. So, equilibrium point ($C, C)$ is an equilibrium which the players selected based on the benefit of the opponent. Therefore, Based on the hyper-rationality concept, prisoner's dilemma has two equilibria points ($C, C)$ and $(D, D)$. Hofstadter claims that for the super-rational it is evident that universal cooperation is the best option \cite{poston2012social, hofstadter2008metamagical}. We state that either cooperation or defection is the best option for the hyper-rational player depending on environmental conditions (competitive, peace, war,...) and the kind of behavior interactive (friend, enemy,...) in game. One can almost be claimed that the cooperation and the defection are two hyper-rational options. On the other hand, when players cannot guarantee their profit, they can optimize the situation by defection. As a result, cooperation is not always the best option.

\begin{figure}
\centering
\begin{tikzpicture}
\node [ opmobject] (4) {\begin{tabular}{c|c|c|}
\multicolumn{1}{c}{$g_{1}$} & \multicolumn{1}{c}{} & \multicolumn{1}{c}{} \\[-2.5mm]
\multicolumn{1}{c}{} & \multicolumn{1}{c}{C} & \multicolumn{1}{c}{D} \\ \cline{2-3}
C & 3,3 & 1,4 \\ \cline{2-3}
D & 4,1 & 2,2 \\ \cline{2-3}
\end{tabular}};
\end{tikzpicture}
\caption{Prisoner's dilemma $g_1$}
\label{fig:PC}
\end{figure}

Cooperation leads to a tension between what is best for the individual and what is best for the group \cite{dreber2008winners}. How can the cooperation be promoted in the repeated prisoner’s dilemma? Some experimental papers giving support to the promotion of cooperation have used from complicated laboratory conditions or add strategy to Prisoner's Dilemma \cite{sigmund2001reward, fehr2002altruistic, ohtsuki2009indirect, traulsen2006evolution, mathew2011punishment, gallo2015effects, jordan2016third, bear2016intuition}. We want to answer the above question using the concept of hyper-rationality. In classical game theory, all the variety of $2\times 2$ symmetric games rests in the relative values of the four payoffs $(C, C)$, $(D, D)$, $(D, C)$ and $(C, D)$. There are twenty-four possible rankings of four payoffs, and thus twenty-four symmetric $2\times 2$ games \cite{poundstone1993prisoner}. In Prisoner’s Dilemma based on classical rationality, row player prefers $(D, C)$ to $(C, C)$ to $(D, D)$ to $(C, D)$ ($(D, C) \succeq (C, C) \succeq (D, D) \succeq (C, D)$). The hyper-rationality of player thinks about profit or loss of other players in addition to his personal profit or loss and then will choose an action which is desirable to him. As mentioned above, taxonomy of player’s hyper-preferences depends on environmental condition, the kind of behavior interactive, self-evaluation system and evaluation system of other interacting persons. Taxonomy of player’s hyper-preferences has four main behavioral options: individual profit, individual loss, profit for others and loss for others. Hence, for row player take taxonomy of hyper-preferences considering four main behavioral options: based on individual profit: $(D, C) \succeq^{'} (C, C) \succeq^{'} (D, D) \succeq^{'} (C, D)$, based on individual loss: $(C, D) \succeq^{'} (D, D) \succeq^{'} (C, C) \succeq^{'} (D, C)$,  based on profit for others: $(C, D) \succeq^{'} (C, C) \succeq^{'} (D, D) \succeq^{'} (D, C)$, based on loss for others: $(D, C) \succeq^{'} (D, D) \succeq^{'} (C, C) \succeq^{'} (C, D)$.

Taxonomy of hyper-preferences helps to players consider two main behavioral options simultaneously, for example: collective profit or collective loss. If interaction between players is based on collective benefit thinking, both player prefers: either $(C, C) \succeq^{'} (D, D) \succeq^{'} (D, C) \succeq^{'} (C, D)$  or  $(C, C) \succeq^{'} (D, D) \succeq^{'} (C, D) \succeq^{'} (D, C)$. On the other hand, the concept of hyper-rationality explains that, based on the profit of other players, cooperation is a strictly dominant action. Taxonomy of hyper-preferences helps to players consider two principles simultaneously in interactions: first, treat others as we ourselves would like to be treated and second, not treat others as we ourselves would not like to be treated. According to the second principle, the profiles strategy $(C, D)$ and $(D, C)$ is not selected. Based on collective benefit thinking $(C, C)$ is preferred to $(D, D)$ by two players. So, according to the first principle, the profiles strategy $(C, C)$ is selected. In repeated Prisoner's Dilemma, the taxonomy of two principles are likely to lead to this mechanism for the promotion of cooperation. Hence, this study with the help of the concept of hyper-rationality in game theory, seeks to the promotion of cooperation between the players, which is a sign of the power of this concept. Therefore, the concept of hyper-rationality helps to understand the evolution of cooperation.

There are two competing classes of theories have been proposed to explain the relationship between \textit{actions} and \textit{beliefs}. The first class of theories, coming from economics, suggests that beliefs cause actions. The second class of theories, proposed by psychologist, suggest that actions cause beliefs \cite{croson2004explaining, attanasi2008survey}. Interactions between psychologists and economists have been marked more by conflict than by collaboration \cite{manski2017collaboration}. Bruine de Bruin and Fischhoff describe four conditions that promoted transdisciplinary collaborations between psychologists and economists \cite{de2017eliciting}. We explore the players' behavior with the help of the concept of hyper-rationality. This concept explains that, based on the loss of other players, defection is a strictly dominant action. If interaction between players is based on collective loss thinking, both player prefers: either $(D, D) \succeq^{'} (C, C) \succeq^{'} (D, C) \succeq^{'} (C, D)$, or $(D, D) \succeq^{'} (C, C) \succeq^{'} (C, D) \succeq^{'} (D, C)$. These interpretations help to enlarge our understanding of psychological aspects of strategy choices in games and also provide an analysis of the decision-making process with cognitive economics approach at the same time. For example, considering two principles above and based on collective thinking, prefers $(D, D)$ to $(C, C)$ by two player shows that players are spiteful individuals and only think to cause loss to others. In other words, the hyper-preferences indicate that the kind of behavior interactive, environmental conditions, and valuation system are based on hostility and players at this point have considered the maximum loss to other, which this is a common research language between psychologists and economists.
\subsection*{Missile crisis game }
In \emph{Theory of Moves,} Steven Brams developed a general dynamic modeling framework and used it to evaluate Cuban missile crisis \cite{brams1994theory}. In this book, he explains that the goal of the United States was immediate removal of the Soviet missiles, and United States policymakers seriously considered two strategies to achieve this end:\\
1. A naval blockade (B), or "quarantine" as it was euphemistically called, to prevent shipment of further missiles, possibly followed by stronger
action to induce the Soviet Union to withdraw those missiles already installed.\\
2. A "surgical" air strike (A) to wipe out the missiles already installed, insofar as possible, perhaps followed by an invasion of the island.

The alternatives open to Soviet policy makers were:\\
1. Withdrawal ( W) of their missiles.\\
2. Maintenance (M) of their missiles.\\
Instead, he begins by considering the payoff matrix given in Fig. \ref{fig:Ms}. In Cuban missile crisis, Brams concludes that the two countries reached a compromise. In other words, in this crisis the compromise outcome or $(3, 3)$ is a Nash equilibrium.
\begin{figure}
\centering
\begin{tikzpicture}
\node [ opmobject] (4) {\begin{tabular}{c|c|c|}
\multicolumn{1}{c}{$g_2$} & \multicolumn{1}{c}{} & \multicolumn{1}{c}{} \\[-2.5mm]
\multicolumn{1}{c}{} & \multicolumn{1}{c}{W} & \multicolumn{1}{c}{M} \\ \cline{2-3}
B & 3,3 & 1,4 \\ \cline{2-3}
A & 2,2 & 4,1 \\ \cline{2-3}
\end{tabular}};
\node [ opmobject, right=of 4, xshift=8 pt,] (8) {\begin{tabular}{c|c|c|}
\multicolumn{1}{c}{$g_{3}$} & \multicolumn{1}{c}{} & \multicolumn{1}{c}{} \\[-2.5mm]
\multicolumn{1}{c}{} & \multicolumn{1}{c}{C} & \multicolumn{1}{c}{D} \\ \cline{2-3}
C & 4,3 & 2,4 \\ \cline{2-3}
D & 3,1 & 2,1 \\ \cline{2-3}
\end{tabular}};
\end{tikzpicture}
\caption{Missile crisis game $g_2$ and Trickery game $g_3$}
\label{fig:Ms}
\end{figure}

This game has two Nash equilibrium $(A, W)$ and $(B, W)$. In game $g_2$, for the US, based on damage for Soviets, $A$ is a strictly dominant action of loss for Soviets and $B$ is a strictly dominated action of loss for Soviets. Why did not the US an air strike that destroys the missiles? Soviet response to this strategy makes pairs of his rational actions. Recently Eshaghi and Askari \cite{Eshaghi2017dynamic} have introduced the definition of \emph{pair of rational actions}. Therefore, pairs of actions $(A, W)$ and $(A, M)$ are rational for Soviet. Bram's write: an air strike that destroys the missiles that the Soviets were maintaining is an "honorable" U.S. action, (its best state) and thwarts the Soviets (their worst state)- $( 4, 1 )$. An air strike that destroys the missiles that the Soviets were withdrawing is a "dishonorable" U.S. action (its next-worst state) and thwarts the Soviets (their next-worst state)- $(2,2)$. According to the concept of hyper-rationality, based on the loss for the US, pair of actions $(A, W)$ is a pair of hyper-rationality. So, a pair of actions $(A, W)$ is a Nash equilibrium. In this game, for the US, based on benefit for Soviets, $B$ is a strictly dominant action of profit for Soviets and $A$ is a strictly dominated action of profit for Soviets. Therefore, pairs of actions $(B, W)$ and $(B, M)$ are rational for Soviet. According to the concept of hyper-rationality, based on benefit for the US, a pair of actions $(B, W)$ is a pair of hyper-rationality. So, a pair of actions $(B, W)$ is a Nash equilibrium. Consequently, the United States has two dominant actions: $B$ based on benefit to Soviets and $A$ based on the loss to Soviets. On the other hand, if interaction between the United States and the  Soviets is based on collective benefit thinking, both player prefers: either $(B, W) \succeq^{'} (A, W) \succeq^{'} (A, M) \succeq^{'} (B, M)$  or  $(B, W) \succeq^{'} (A, W) \succeq^{'} (B, M) \succeq^{'} (A, M)$. Therefore, in Missile crisis game,  it can be said that the two countries considered the collective benefits and reached a compromise.

\subsection*{Trickery game}
Recently Eshaghi and Askari introduced Trickery game \cite{Eshaghi2017dynamic}. This game shows a good reason for the trickery of some people in everyday life, which one of the players can with cunning change his action that reduces the payoff, other players. We consider row player as player 1 and column player as player 2. The trickery game $g_3$ table is given in Fig. \ref{fig:Ms}. The above game is an asymmetric game. This game has three Nash equilibriums $(C, D)$, $(D, D)$ and $(C, C)$. In the trickery game, player 1 has weakly dominant action $C$ and weakly dominated action $D$ based on individual benefit. Player 2 has weakly dominant action $D$ and weakly dominated action $C$ based on individual benefit. Also, in this game, player 1 has strictly dominant action $D$ and strictly dominated action $C$ based on loss of player 2. Player 2 has strictly dominant action $D$ and strictly dominated action $C$ based on loss of player 1. According to the concept of classic rationality, two Nash equilibria $(C, D)$ and $(D, D)$ have equal payoff for player 1, but considering the concept of hyper-rationality, based on loss for player 2 given, $(D, C) \sim^{'} (D, D) \succeq^{'} (C, C) \succeq^{'} (C, D)$, player 1 prefer $(D, D)$ to $(C, D)$. With assumption choosing strictly dominant action $D$ for player 2 and hyper-rationality, concept of hyper-rationality for player 1 rules that if he seeks to benefit player 2, he will choose $C$, so the equilibrium of game is $(C, D)$, if he seeks to cause loss to player 2, he will choose $D$ because the opponent will sustain more loss and his outcome will remain fixed, therefore the equilibrium game is $(D, D)$. Therefore, the concept of hyper-rationality helps to understand a common research language between psychologists and economists.

In this game, there are two Nash equilibrium points and based on the definition of Nash equilibrium, no player will be motivated to change his action in equilibrium point. Based on concept of hyper-rationality, given fixed defect of player 2, we can see that if layer 1, seeks to incur loss to his opponent, he will be motivated to change his action in equilibrium point $(C, D)$ and can transfer game equilibrium to point $(D, D)$. We want to see what reaction player 2,  will show if player 1, chooses $D$. With choosing $D$ by player 1, choosing $C$ and $D$ will give equal outcome for player 2, but if player 2, prefer loss of opponent, we will find that he doesn’t choose strictly dominated action $C$ and by choosing strictly dominant action $D$ will incur more loss to player 1. So, we conclude that pair of action $(D, C)$ is not chosen and player 2, is not motivated to change his action in equilibrium point $(D, D)$. Therefore, action profile $(D, D)$ is a hyper-equilibrium. In Trickery game, player 1, can use this hyper-equilibrium as a credible threat, to force player 2 to choose cooperation. So, action profile $(C, C)$ is an equilibrium which chooses based on the collective benefit and pressure of the player 1.


\section*{Discussion}
The theory of rational choice seeks to explain the behavior of the person who behaves wisely and look for to maximize his benefits. The theory of hyper-rational choice seeks to explain the behavior of the person who behaves wisely and considers benefit or loss of others in addition to the individual benefit. The theory of hyper-rational choice can be applied instead of the theory of rational choice in social sciences and society can be analyzed based on two methodological individualistic and methodological collective approaches so that human society can be understood easily and predicted more fluently. In this theory, the hyper-rational actor thinks about the benefit (collective benefit) and loss (collective loss) of other actors beside his own benefit and then chooses an action. We also presented a new definition as hyper-rationality. Based on this concept, we call a person hyper-rational if results of choosing any action are comparable with other actions for the person and also the person can recognize what action is the most beneficial to him among the accessible actions and what action causes loss or benefit of other players. Based on the concept of hyper-rationality, a player may not recognize that what action is the most beneficial to him but can choose an action which causes the maximum loss or benefit for other players. According to this concept, assumption about profit or loss to other actors can be added in the form of human behaviors such as altruism, devotion, fraud, jealousy, and mistrust to the concept of rationality. This new concept can describe some of the human behaviors well.

Moreover, new definitions such as the best response functions, strictly dominant action and weakly dominant action based on loss of other players. With the help of these definitions, we have introduced a method for finding equilibrium in the game and classified the equilibriums. In some of these games, we achieved equilibriums which are not considered as game equilibrium based on the classic concept of rationality. We presented an important game as trickery game which has special equilibrium which we call hyper-equilibrium. The hyper-equilibrium is the point in which only one player can displace equilibrium to another point by changing his action which causes profit or loss to other players so they cannot change their action. The theory of hyper-rational choice can be applied as a basis for the theory of decision-making, the theory of collective action and social sciences issues and new results can be obtained in this way. Hyper-rationality can be mentioned and investigated as uncertainty. We also believe that our current design has some additional advantages over previous ones


\end{document}